\documentclass[conference]{IEEEtran}
\IEEEoverridecommandlockouts
\usepackage{cite}
\usepackage{amsmath,amssymb,amsfonts}
\usepackage{algorithmic}
\usepackage{graphicx}
\usepackage{textcomp}
\usepackage{xcolor}
\usepackage{booktabs}
\usepackage{balance}
\usepackage{multirow}
\usepackage{color}
\usepackage{hyperref}

\usepackage{adjustbox} % For fine control
\usepackage{amssymb}
\usepackage{tikz}
\newcommand*\circled[1]{\tikz[baseline=(char.base)]{
            \node[shape=circle,fill,inner sep=1pt] (char) {\textcolor{white}{#1}};}}

\hypersetup{
    colorlinks=true,
    linkcolor=blue,
    filecolor=magenta,      
    urlcolor=cyan,
    citecolor = magenta,
    pdfpagemode=FullScreen,
}

\begin{document}

\title{\vspace{-20pt}HyDra: SOT-CAM Based Vector Symbolic Macro for Hyperdimensional Computing \vspace{-15pt}}
\author{
    \IEEEauthorblockN{
        Md Mizanur Rahaman Nayan, 
        Che-Kai Liu, 
        Zishen Wan, 
        Arijit Raychowdhury,
        Azad J Naeemi
    }
    \IEEEauthorblockA{
       Department of Electrical and Computer Engineering,\\ 
        Georgia Institute of Technology, USA
        %\\
        %Email: \{mnayan6, ritik.raj, gshaik6, tushar, azad\}@gatech.edu
    }
    \vspace{-40pt}
}

\maketitle

\begin{abstract}
Hyperdimensional computing (HDC) is a brain-inspired paradigm valued for its noise robustness, parallelism, energy efficiency, and low computational overhead. Hardware accelerators are being explored to further enhance their performance, but current solutions are often limited by application specificity and the latency of encoding and similarity search. This paper presents a generalized, reconfigurable on-chip training and inference architecture for HDC, utilizing spin-orbit-torque magnetic random access memory (SOT-MRAM) based content-addressable memory (SOT-CAM). The proposed SOT-CAM array integrates storage and computation, enabling in-memory execution of key HDC operations: binding (bitwise multiplication), permutation (bit shfiting), and efficient similarity search. Furthermore, a novel bit drop method-based permutation backed by holographic information representation of HDC is proposed which replaces conventional permutation execution in hardware resulting in a 6× latency improvement, and an HDC-specific adder reduces energy and area by 1.51× and 1.43×, respectively. To mitigate the parasitic effect of interconnects in the similarity search, a four-stage voltage scaling scheme has been proposed to ensure an accurate representation of the Hamming distance. Benchmarked at 7nm, the architecture achieves energy reductions of 21.5×, 552.74×, 1.45×, and 282.57× for addition, permutation, multiplication, and search operations, respectively, compared to CMOS-based HDC. Against state-of-the-art HDC accelerators, it achieves a 2.27× lower energy consumption and outperforms CPU and eGPU implementations by 2702× and 23161×, respectively, with less than 3\% drop in accuracy.
\end{abstract}

\section{Introduction}
\label{sec:intro}
Artificial intelligence (AI) is greatly influencing the current and future of computation. However, the state-of-the-art (SOTA) AI systems are power and resource hungry because of which SOTA AI applications are not readily deployable on edge devices. To overcome this challenge, researchers are looking at novel computing paradigms, especially for edge devices\cite{khaleghi2021tiny}.  In this context, Hyperdimensional computing (HDC) has emerged as an alternative with major advantages in terms of energy efficiency, robustness to noise,  and resource requirements \cite{kleyko2023survey}. HDC has already outperformed conventional algorithms in numerous simple tasks such as DNA pattern matching, clustering, digit recognition, activity recognition, and speech recognition~\cite{hernandez2021framework,wang2022odhd,ibrahim2024special,zhang2022scalehd}. Beyond AI applications, HDC has shown great potential in other domains such as robotics for cognitive reasoning~\cite{ibrahim2024efficient, hersche2023neuro,langenegger2023memory, menon2022role,wan2024h3dfact}. Such a broad range of potential applications has motivated researchers to introduce hardware accelerators for HDC \cite{khaleghi2021tiny,ibrahim2024efficient, khaleghi2022generic,shou2023see}. However, these platforms suffer from significant throughput limitations due to the lengthy processes involved in encoding and similarity search~\cite{chang2023recent}.

To speed up the search operation in HDC, CAMs have been proposed. In particular, CAMs based on emerging memory devices have attracted attention for edge applications thanks to their non-volatility. However, in existing proposals, other key HDC operations are performed outside of memory~\cite{langenegger2023memory,chang2023recent}, which limits both throughput and energy efficiency. Enabling in-memory execution of a broader set of HDC operations holds significant promise for improving the overall performance and energy efficiency of these accelerators..

%%In this work, we introduce the implementation of the key HDC operations such as binding, similarity search and permutation within the unified CAM blocks by redesigning the CAM cells and arrays. 

In this work, we demonstrate execution of the key HDC operations such as binding, permutation and similarity search within the SOT-CAM arrays by introducing SOT-CAM-based vector symbolic architecture along with multi-stack design exploitation. Among the emerging memories, we focus on SOT memory because of its ultra-low leakage power, fast write, high density, non-volatility, low voltage, and compatibility with CMOS processes\cite{worledge2024spin}. Moreover, we perform all fundamental HDC operations within the CAM arrays because of which the proposed platform is general purpose in contrast to prior approaches, which focused on specific applications that required only a subset of HDC operations. This feature enables mapping various HDC applications like classifications and clustering. The proposed design is also reconfigurable and support various hypervector (HV) dimensions by reconfiguring CAM banks which offers higher energy efficiency without compromising performance for a specific application.

The key innovations of the proposed vector symbolic macro for HDC applications capable of executing both search and encoding within the CAM array are:

% \CL{
% \begin{itemize}
%     \item A reconfigurable SOT-MRAM based cell that unifies in-memory computation and content-addressable functionality.
%     \item Algorithmic-driven permutation acceleration provides xxx
%     \item IR-drop mitigation search scheme to counter advanced node interconnect parasitics.
% \end{itemize}
% }

% \begin{itemize}
%     \item An SOT-CAM based HDC architecture which has a reconfigurable unified storage and computation unit for item hypervector (HV), level HV and class HV, capable for both training and inference.
%     \item A modified CAM cell design along with the array that performs the XOR operation between the stored and query, resulting in the acceleration of bipolar HV multiplication. It also enables the cell to function as both conventional and associative memories.
%     \item Demonstration that 8- or 16-bit shift operations can replace single bit shift operations which enables simple modification of the read operation to perform permutations thanks to holographic information representation of HDC.
%     \item Voltage scaling to improve search non-ideality posed by interconnect parasitic during in-array similarity search through current sensing.
%     \item  Simplified 16-bit adder design that improves area and energy footprint by $1.43\times$  and $1.51\times$,  respectively, over a conventional 16-bit adder. \CL{How}

% \end{itemize}

\begin{itemize}
    \item \textbf{Cell.} A reconfigurable $5T2MTJ$ SOT-CAM cell that unifies in-memory execution of key HDC operations with storage. 
    \item  \textbf{Array.} A voltage scaling scheme to mitigate IR drop impact on accurate distance representation during search.
    \item \textbf{Architecture.} Supports all key HDC operations and various HV dimensions, enabling the execution of a wide range of applications and allowing targeted optimization for energy, performance, and latency.
    \item \textbf{Algorithm-Hardware Co-Design.} Leveraging holographic information representation to enable in-memory execution of permutation operation.
\end{itemize}

The rest of the paper is organized as follows. Section~\ref{sec:background} presents background on HDC, including key operations, application pipeline, and limitations of existing hardware platforms, motivating the use of SOT-CAM as a compute-in-memory unit. Section~\ref{sec:hydra} introduces HyDra, our SOT-CAM-based architecture, detailing its design methodology, cell and array implementation, algorithm-hardware co-design, and voltage scaling for accurate distance computation. Section~\ref{sec:implementation} presents implementation and evaluation results, validating key design choices and demonstrating improvements in performance, accuracy, and energy efficiency. Section~\ref{sec:conclusion} concludes the paper.

\begin{figure}
    \centering
    \vspace{-5pt}
    \includegraphics[width=\linewidth]{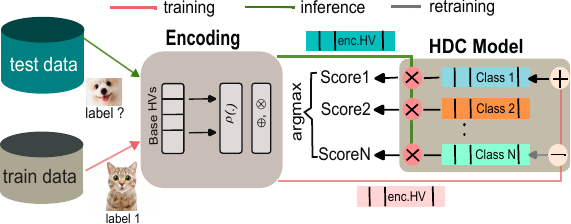}
    \vspace{-15pt}
    \caption{Simplified dataflow of HDC model training, retraining and inference. During training encoded HV are added to corresponding class HV. In retraining, encoded HV are subtracted from the mispredicted class HV and added to correct class HV. In inference, most similar class is given as prediction. }
    \vspace{-15pt}
    \label{fig:HDpipeline}
\end{figure}

% \vspace{-1pt}
\section{Background and Related Work}
\label{sec:background}
\subsection{Key HDC Operations}
In HDC, encoding refers to mapping information in hyperspace in the form of HVs  and decoding is to extract decision from the represented information in hyperspace. Among the many approaches for encoding, the Multiply-Add-Permute (MAP) is the most popular\cite{9921397}.  In MAP, elementwise multiplication (binding), elementwise addition (bundling) and permutation (bit shifting) are the three fundamental operations. Multiplication and permutation operations, performed on the binary HVs representing input symbols and levels, result in binary outputs. We refer to these as binary HDC operations. The results of the addition operation on the other hand can become large integers; hence, they cannot be represented with binary elements without losing information. This is why we call addition a multibit HDC operation.  

\subsection{Applications Pipeline} 
\textbf{Classification.} The dataflow of HDC (Fig. \ref{fig:HDpipeline}) during training and inference can be divided into two parts: encoding and similarity search. During encoding, inputs are converted to HVs. During training, the HVs associated with a subset of inputs are summed up to create class HVs. For inference, the HV of an input is compared with the class HVs to find the closest match.

\textbf{Clustering.} K-means like clustering can be executed in hyperspace as presented in \cite{imani2019hdcluster}. During clustering, data points to be clustered are encoded and stored for clustering. Next, random HVs are initialized for the K cluster centers. In each epoch, data point HVs are assigned to clusters based on similarity with the HVs of the cluster centers. At the end of every epoch, the HV's corresponding to the cluster center can be updated by adding data point HV's of a cluster and performing the sign operation to binarize until the similarity distance (e.g., Hamming distance) between the previous and new cluster center is lower than a threshold. 

% \vspace{-10pt}

\subsection{Existing HDC Computing Platforms and Their Limitations}

Prior efforts to develop hardware platforms for HDC can be classified as microprocessor-based systems, FPGA implementations, and compute-in-memory (CIM) architectures.  

Microprocessor-based embedded systems target ultra-low-power IoT applications by optimizing dataflow and memory use to accommodate small cache sizes. Challenges include achieving bit-level parallelism, scalability, and reconfigurability. PULP-HDC \cite{montagna2018pulp} introduced a low-power parallel architecture utilizing tightly coupled data memory (TCDM) as a scratchpad among RISC cores. Tiny-HD~\cite{khaleghi2021tiny} lowered the base HV memory requirements using circular rotation and matrix-vector encoding. However, the restricted hardware used for random projection encoding requires usage of multibit data representation, and hence large multipliers~\cite{chang2023recent}. 

FPGA-based platforms leverage HDC's Boolean HVs and parallelism. Despite their programmability\cite{salamat2019f5, salamat2020accelerating, schmuck2019hardware}, they suffer from computationally intensive encoding, expensive associative searches, and high leakage in lookup tables~\cite{chang2023recent}.  

CIM architectures integrate computation within memory to reduce data movement and energy use, aligning well with HDC's parallelism and noise resilience. Examples include HAM \cite{imani2017exploring} for analog associative search, 3D RRAM crossbars, and PCM-based neural networks for few-shot learning \cite{li2016hyperdimensional, karunaratne2021robust}. Optimizations include PCM-based associative searches \cite{karunaratne2020memory}, DUAL for clustering~\cite{imani2020dual}, and MimHDC for multi-bit HDC \cite{kazemi2021mimhd}. Recent designs like BioHDC \cite{zou2022biohd} enhance sequence search efficiency by 116.1× over GPUs. These architectures often hardwire specific encoding schemes, limiting generalizability. In contrast, this work proposes SOT-CAM arrays to implement core HDC operations, enabling diverse HDC applications with improved throughput and energy efficiency.

\begin{figure}
    \centering
    \includegraphics[width=\linewidth]{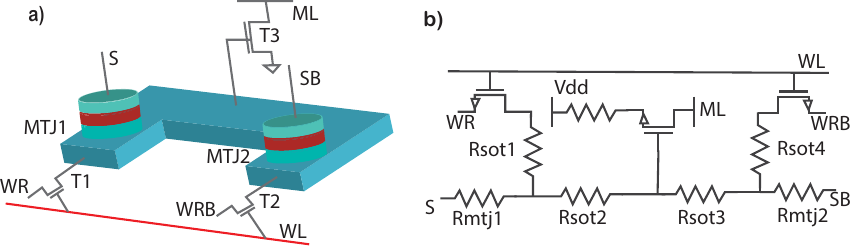}
    \vspace{-10pt}
    \caption{a) 3T2MTJ SOT-CAM cell. b) Equivalent circuit}
    \vspace{-15pt}
    \label{fig:sotcamcell}
\end{figure}
\subsection{SOT-CAM as Emerging Memory and Compute Unit}
Fig.~\ref{fig:sotcamcell} illustrates SOT-MRAM $3T2MTJ$ based SOT-CAM cell where MTJs store complementary data through spin orientation. During the search operation in a cell which is a simple XOR operation between the stored bit and query (applied at the S node), energy consumption is negligible due to the very high resistance of the MTJs (in $M\Omega$ order). Apart from being non-volatile, SOT-CAM has other advantages. It has high endurance of $10^{12}$, low read error rate due to complimentary data read, and low write error rate of $10^{-6}$ enabled by reduced SOT current by leveraging voltage-controlled magnetic anisotropy (VCMA) effect where a voltage is applied through the search (S) node\cite{nguyen2024recent, narla2022design}. Note that the U shape of the SOT layer ensures complementary write-through current flowing in opposite directions beneath the two MTJs. In terms of latency, SOT-CAM achieves read, write, and search latency similar to SRAM while providing a higher density than SRAM\cite{narla2022design}. Additionally, wafer-level demonstration justifies technology maturity and CMOS compatibility\cite{yasin2024extremely}.

\section{HyDra: Design and Methodology}
\label{sec:hydra}
%%\vspace{-1pt}
In this section, we present the HyDra architecture and its supporting methodology. We begin by detailing the cell and array design through an illustrative example followed by multiplication and similarity search mapping on SOT-CAM array. Next, we demonstrate the algorithm-hardware co-design approach, highlighting the proposed permutation scheme and an efficient adder design. Finally, we discuss a voltage scaling techniques to mitigate the impact of IR drop on search performance.

\subsection{SOT-CAM-Based Architecture}

\begin{figure}[tb]
    \centering
    \vspace{-10pt}
    \includegraphics[width=\linewidth]{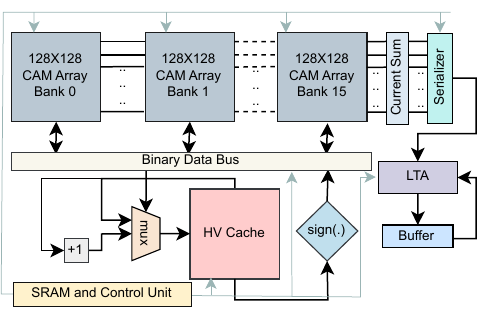}
    \vspace{-15pt}
    \caption{Proposed HyDra architecture with SOT-CAM.}
    \label{fig:HPU Architecture}
    \vspace{-15pt}
\end{figure}

Fig.~\ref{fig:HPU Architecture} represents the proposed architecture of the HDC accelerator. 16 SOT-CAM array banks function as both storage and computing units. Each bank consists of a 128$\times$128 array of size $16Kb$. To ensure reconfigurability, and energy efficiency, the entire CAM block has been split into CAM banks. They are used for storing basis and level HVs of various dimensions.  During encoding, binding and permutation operations are executed in the arrays.  Similarity search based on the Hamming distance is also performed in the arrays during inference. 

To ensure no information is lost during encoding, the HV cache block is used to temporarily store the HV in the $int16$ data type. The corresponding HV of a feature or datapoint is updated very often. Thereby, caching them improves the speed and overall energy consumption. After encoding or training, the cache blocks data with $int16$ precision are binarized to be stored in the CAM array for further usage. A current sum block is used to add match-line (ML) currents of each bank corresponding to each row. This block selects active banks which depend on the HV dimension. A serializer is used to pass current from one batch of rows to Loser Takes All (LTA) block. This approach adds further programmability to work with multiple classes along with adding reusability of the same block results in lower area overhead. LTA blocks find the candidate with the lowest current and then store it in a buffer. This candidate is used again in the next stage to be used for selecting the most similar candidate (lowest ML current) for the final prediction. 

Note that the proposed architecture supports all the HDC operations. Additionally, the architecture support of different HV dimensions enables optimization of energy consumption and system throughput, while maintaining acceptable performance levels that can vary significantly across datasets.

\begin{figure*}
    \centering
    % \vspace{-15pt}
    \includegraphics[width=\textwidth]{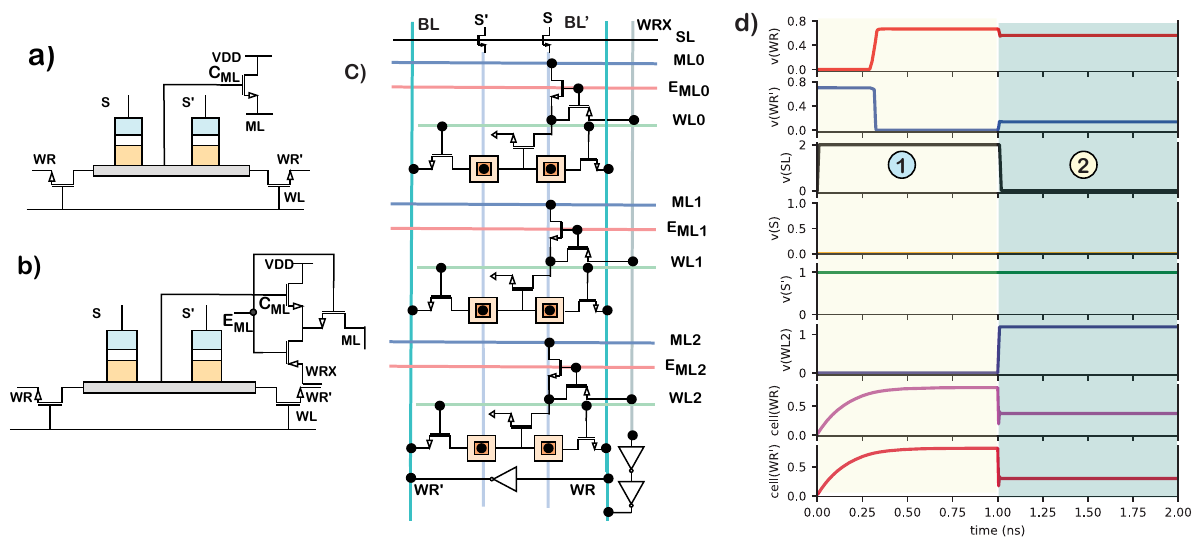}
    \vspace{-25pt}
    \caption{a) Simple SOT-CAM cell design. b) The proposed $5T2MTJ$ SOT-CAM cell design for the HDC array. c) Proposed array level design (one column).  PMOS connected to the $E_{ML}$ propagates the XOR output to the write driver, $WRX$. And NMOS connects the cell to the corresponding ML during search. d) Output waveform during performing an XOR-Write operation. First phase, XOR operation is performed by enabling search line (i.e., $V(SL)$ is high for first 1ns).  Target WL(e.g. $WL2$) is turned on in second phase to write the XOR output. Voltage difference between $WR$ and $WR'$ is $~78mV$ that result in sufficient SOT current ($156uA$) to write with certainty.}
    \vspace{-10pt}
    \label{fig:cellArrayXOR}
\end{figure*}

% \vspace{-5pt}

\subsection{Cell and Array Design for HDC Operation}
SOT-CAM arrays are the major computational block of the proposed architecture since they performs most of the fundamental operations during encoding and perform the similarity search during inference. Fig.~\ref{fig:cellArrayXOR}(a) and (b) depict the cell design used in the array. 

\subsubsection{\textbf{Simple SOT-CAM cell design}} Fig.~\ref{fig:cellArrayXOR}(a) shows a simple CAM cell with the voltage at node $C_{ML}$ being at high (low) when there is a mismatch (match) between the cell's stored value and the search bit. $C_{ML}$ drives the NMOS connected to the ML that is shared among the cells on the same row. Note that, that complementary values are stored in the two MTJs and complementary search voltages on search lines are applied. Voltage division happens across SOT layer which generates high or low voltage at $C_{ML}$ node\cite{narla2022design}. During similarity search, the currents from cells connected to the ML get added and the sum reflects the Hamming distance between the stored vector and the query. An LTA block is used to find the smallest current that represents the most similar candidate (prediction). During write operation, WL is activated and bit line is connected to WR and WR' which inject current through the SOT layer to align spin in the MTJ according to bit line. 

\subsubsection{\textbf{Proposed 5T2MTJ cell and array design}} Fig.~\ref{fig:cellArrayXOR}(b) depicts the proposed 5T2MTJ cell that can perform multiplication and similarity search which can also be used as a random access memory. The additional two transistors are to equip the cell with reconfigurable functionalities for multiplication and similarity search operation. $E_{ML}$ signal is used to configure the cell by either connecting it to ML during similarity search or to the $WRX$ node during multiplication or simple memory read. The transistors isolate the cell from the ML during multiplication and isolate it from WRX node during similarity search to avoid adjacent disturbances. The column-level array design is depicted in Fig.~\ref{fig:cellArrayXOR}(c) where one vertical interconnect is connected to each cell's WRX node. 

\subsubsection{\textbf{Single bit XOR operation}}Fig.~\ref{fig:cellArrayXOR}(d) demonstrates performing the XOR operation where the operation has been conducted between the cell in the first row and the vector element on the search line. The XOR output has been written to the third row.  \circled{1} In the first phase of the operation, XOR results are made available to the write drivers ($WR$ and $WR'$) through $WRX$. This takes around $270ps$ to get the stable output in the write driver. \circled{2} In the second phase, the output of the XOR operation is written to the target row (third row in this case). $(WL2)$ is turned on after $1ns$. To ensure a reliable write operation, we need to make sure enough SOT current is being delivered by the write driver which is about $140uA$. This current ensures spin switching probability to 1, resulting in a successful write operation \cite{narla2022design}. Note that the voltage difference between WR and WR' is about $78mV$, which results in $156uA$ SOT current.

% \begin{figure}
%     \centering
%     \vspace{-5pt}
%     \includegraphics[width=\linewidth]{figures/adderHDC.pdf}
%     \vspace{-15pt}
%     \caption{Simple incrementor block for addition operation. Operand A is only incremented by 1 when HV element to be added is 1 otherwise unaffected. }
%     \vspace{-15pt}
%     \label{fig:HDCadder}
% \end{figure}

\begin{figure*}
    \centering
    \includegraphics[width=\textwidth]{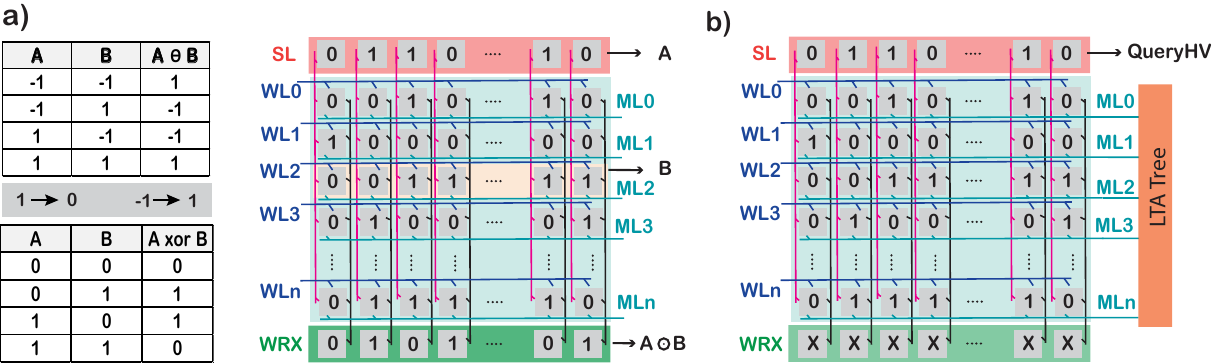}
    \caption{a) Multiplication mapping on SOT-CAM array. One operand HV is loaded in SL where another one is in the array, and elementwise output is passed through $WRX$ node. b) Similarity search mapping. Query HV is applied in $SL$ and candidate HVs are inside the CAM arrays. Similarity search between query and all candidate HVs in the array is fully parallel where the Hamming distances are reflected in the corresponding ML current.}
    \vspace{-10pt}
    \label{fig:multSearchMap}
\end{figure*}

\subsubsection{\textbf{Multiplication operation}}Multiplication among bipolar basis and level HV always generates bipolar HV. We can map this to XOR operation where 1 and -1 are mapped to 0 and 1, respectively (Fig.~\ref{fig:multSearchMap}(a)). To perform an XOR operation on two HV's, one HV is loaded on the search lines (S and S'). The row containing the other HV is activated through  $E_{ML}$ that connects WRX and isolates the cells on the row from $ML$.

\subsubsection{\textbf{Similarity search operation}} Each CAM bank of 128$\times$128 performs as an individual search unit which generates ML currents in each row of the array. During similarity search, $E_{ML}$ connects the output of the CAM cell to the ML by activating its NMOS and isolates it from the $WRX$ through its PMOS. Query HV is loaded in the SL and similarity search is performed against all the HVs inside the array (Fig.~\ref{fig:multSearchMap}(b)). Ideally, each cell should source an identical amount of current into ML during a mismatch between the stored and query bits to ensure a linear relationship between the ML current and the Hamming distance between the corresponding portions of the stored and query vectors.  The output currents in each CAM banks are summed across the rows to get the total current representing the Hamming distance between the stored and query vectors. To reduce the hardware overhead, we use 8-input LTA block for comparison and determine most similar class (Fig.~\ref{fig:HPU Architecture}). We split all the currents corresponding to classes in batches to be compared in the LTA block. At a time, the LTA block takes one batch of 8 ML currents and finds the lowest one index, which is stored in the buffer and then the next batch of 8 ML current including the last selected current and new 7 ML currents. Thus, we can leverage the same hardware to make the final decision.

\subsubsection{\textbf{Conventional memory read in CAM}} In HyDra, CAM blocks offer storage of HV apart from being used as computing modules. This objective requires HV to be accessed by index similar to conventional memory. In the proposed design, target row's WL needs to be activated and 0 needs to apply to all search lines. This mechanism is simply based on: [$X$ \textbf{xor} $0$] generates X in the output node WRX.  In the proposed cell (Fig. \ref{fig:cellArrayXOR}(b)), placing a 0 on all the S nodes allows the XOR output at WRX to directly reflect the value stored in the CAM cell within the digital domain. Note that, the output is digital and ready to write in the target address. This approach facilitates straightforward data transfer or simple memory access without expensive sense amplifiers. During the XOR operation, ML is disconnected through $E_{ML}$ signal which deactivates the NMOS transistor connected to ML resulting in the isolation of the cell from ML and activates the PMOS transistor connected to a pre-discharged WRX to drive it to Vdd if the stored value is '0'. Otherwise, the pre-discharged WRX remains low. With this, a simple inverter is used to read the state of the memory bit. The result can be used to write back to any activated row through two other inverters that drive WR and WR'.

\begin{figure}
    \centering
    % \vspace{-10pt}
    \includegraphics[width=\linewidth]{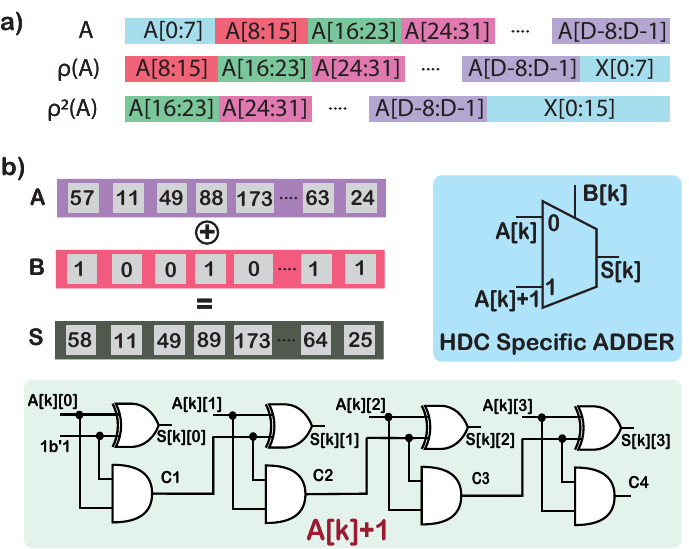}
    \caption{a) Proposed HDC algorithm inspired permutation technique. b) HDC specific adder design. Operand A is only incremented by 1 when HV element to be added is 1 otherwise unaffected.}
    \label{fig:permuteAdder}
    \vspace{-10pt}
\end{figure}

\subsection{Algorithm-Hardware Co-Design}
\subsubsection{\textbf{Holographic information representation in HV and permutation in HyDra}}
Permutation is used to encode temporal or sequential information of data in HDC. In prior HDC hardware designs, bit shifter is used to perform the permutation operation which requires data transfer between memory to compute block. HyDra removes the requirement of using expensive bit shifters and memory to compute unit transfer latency by exploiting holographic information representation of HDC. In hyperspace, information is distributed across all orthogonal dimensions in contrast to the traditional vector space where all dimensions are needed for accurate information retrieval. Thus, even if a portion of a HV is lost, its information still remains preserved in other dimensions.  The proposed permutation technique leverages this property by read the HV operand and storing the HV in the target address by amount we want to shift the HV (Fig.~\ref{fig:permuteAdder}(a)). The end of the HVs is compensated by random initialization which doesn't impact performance, thanks to the holographic information representation of HVs. This approach facilitates straightforward data transfer to other rows or enables permutation functionality. In the design, 16-to-1 multiplexers are employed to select batches, ensuring efficient permutation during read operations. This follows the conventional memory design approach, where sense amplifiers are shared across columns to reduce the cost of having a dedicated sense amplifier for each column.

\subsubsection{\textbf{Addition operation}}
During encoding of a data sample into HV space, each feature of the sample is encoded in HV by binding and permutation of basis and level HV. Then, the resulting HV is bundled with the corresponding class HV. During the bundling operation, the feature HV remains in the 16-bit integer format to employ information where the new feature encoded by binding and permutation is bipolar (Fig.~\ref{fig:permuteAdder}(b)). This enables to implement the elementwise addition operation with  half adder based implementation instead of conventional full adder implementation.  This helps to design a low power adder for HDC as depicted in Fig.~\ref{fig:permuteAdder}(b) where the incremented value or the unchanged element is passed through the MUX if binary operand is 1 or 0 (used to configure MUX), respectively. 

\begin{figure}
    \centering
    % \vspace{-5pt}
    \includegraphics[width=\linewidth]{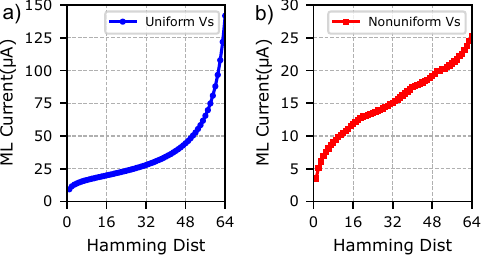}
    \vspace{-15pt}
    \caption{(a) ML current vs Hamming distance non-linearity due to interconnects parasitic. This is caused by IR drop across interconnects that decrease the $V_{GS}$ of the driver NMOS, thus reducing the current contribution by the cell driver. Non-uniform search voltage can improve the non-linearity effect by changing mismatch voltage of the cell. (b) The impact of proposed search voltage scaling where ML current becomes mostly linear.}
     % \CL{Do you mean there are 4 different values being used to obtain this fig?}
    \vspace{-10pt}
    \label{fig:voltageScaling}
\end{figure}
\vspace{-5pt}

% \begin{table*}[ht]
% \vspace{-5pt}
% \caption{Comparison of HyDra (SOT-CAM array) and All CMOS implementation}
% \vspace{-10pt}
% \centering
% \begin{tabular}{|c|c|c|c|c|c|c|c|c|c|}
% \hline
% \multirow{2}{*}{\textbf{Ops}} & \multicolumn{4}{c|}{\textbf{HyDra (SOT-CAM array)}} & \multicolumn{5}{c|}{\textbf{All CMOS (Verilog),  Cycle Time = 0.5ns}} \\
% \cline{2-10}
%  & \textbf{Latency (ns)} & \textbf{Energy (pJ)} & \textbf{Area} & \textbf{In MEM?} & \textbf{Latency} & \textbf{Energy (pJ)} & \textbf{Net Energy (pJ)} & \textbf{Area (mm$^2$)} & \textbf{In MEM?} \\
% \hline
% Addition & 0.462 & 41.08 & 0.01615 & x & 385 Cycle & 61.9 & 883.9 & 0.02304 & x \\
% \hline
% Permutation & 15.36 & 0.752 & * & \checkmark &  193 Cycle & 4.66 & 415.66 & 0.003124 & x \\
% \hline
% Multiplication & 1.548 & 569 & * & \checkmark & 385 Cycle & 3.235 & 828.47 & 0.001138 & x \\
% \hline
% Search & 0.985 & 14.65 & * & \checkmark & 1922 Cycle & 29.65 & 4139.7 & 0.006066 & x \\
% \hline
% \end{tabular}
%     \vspace{-10pt}
% \label{Table:opComparison}
% \footnotesize{
% $^*$ Executed in memory so don't incur additional area. 
% }
% \end{table*}

\begin{table*}[ht]
\centering
% \vspace{-10pt}
\caption{Comparison of HyDra and All CMOS implementation}
\vspace{-6pt}
\begin{adjustbox}{max width=\textwidth}
\begin{tabular}{@{} cccccccccc @{}}
\toprule
\multirow{2}{*}{\textbf{Ops}} & \multicolumn{4}{c}{\textbf{HyDra (Based on SOT-CAM array)}} & \multicolumn{5}{c}{\textbf{All CMOS (Verilog), Cycle Time = 0.5ns}} \\
\cmidrule(lr){2-5} \cmidrule(lr){6-10}
 & \textbf{Latency (ns)} & \textbf{Energy (pJ)} & \textbf{Area (mm$^2$)} & \textbf{In MEM?} & \textbf{Latency} & \textbf{Energy (pJ)} & \textbf{Net Energy$\dagger$ (pJ)} & \textbf{Area (mm$^2$)} & \textbf{In MEM?} \\
\midrule
Addition     & 0.462 & 41.08  & 0.01615  & $\times$ & 385 Cycle  & 61.9  & 883.9   & 0.02304  & $\times$ \\
Permutation  & 15.36 & 0.752  & *        & \checkmark & 193 Cycle  & 4.66  & 415.66  & 0.003124 & $\times$ \\
Multiplication & 1.548 & 569   & *        & \checkmark & 385 Cycle  & 3.235 & 828.47  & 0.001138 & $\times$ \\
Search       & 0.985 & 14.65  & *        & \checkmark & 1922 Cycle & 29.65 & 4139.7  & 0.006066 & $\times$ \\
\bottomrule
\vspace{-7pt}
\end{tabular}
\end{adjustbox}

\footnotesize $^*$ Executed in memory so does not incur additional area. $\dagger$ Considers overhead for off-memory compute.
\vspace{-5pt}
% \caption*{\footnotesize $^*$ Executed in memory so does not incur additional area. $\dagger$ Considers overhead for off-memory compute.}
\label{Table:opComparison}
\end{table*}

\begin{figure}[t!]
    \centering
    \vspace{-5pt}
    \includegraphics[width=.85\linewidth]{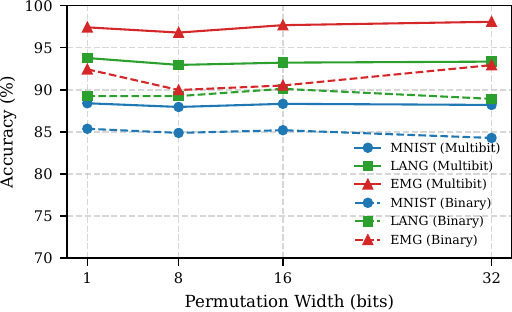}
    \vspace{-5pt}
    \caption{Performance sensitivity on binary HV and multibit HV on three different applications. Binary HV affects performance by 3\% on average, whereas permutation width has a negligible impact on the performance.  }
    \vspace{-10pt}
    \label{fig:Perf}
\end{figure}

\subsection{Search Voltage Scaling for Distance Representation}

\textbf{Impact of IR drop on ML current.} To evaluate the similarity based on the Hamming distance using the ML current, a linear relationship between the ML current and the Hamming distance is essential. Ideally, in our designed array each bit mismatch between the query and stored HV should contribute an equal amount of current to the ML, resulting in a linear relationship between the Hamming distance (i.e., the number of mismatches) and the ML current. However, in practice, the resistance associated with interconnects disrupts this linear relationship. The voltage drop along interconnects introduces variability in the gate-to-source voltage of the driver NMOS transistors in the cells, leading to nonlinearity in the ML current (Fig.~\ref{fig:voltageScaling}(a)). 

\textbf{Significance for overall performance.} The linearity of the readout current is crucial for accurate sensing for several reasons. First, current sensing is highly susceptible to process, voltage, and temperature (PVT) variations. As demonstrated in Fig.~\ref{fig:voltageScaling}(a), non-linear behavior can significantly compromise the accuracy of current-sensing readout circuits, such as the LTA circuitry used in this work, particularly when the measured distance (x-axis) is small. This non-linearity exacerbates sensing errors in such scenarios. Second, similar to voltage sensing, where the input common-mode range (ICMR) governs the allowable range of input signals, the input signals for current sensing must also remain within a specified operating range~\cite{liu2022cosime}. When the distance increases beyond 48 bits, the ML current increases rapidly (Fig.~\ref{fig:voltageScaling}(a)), potentially pushing the current beyond the operating range and causing inaccurate readings. Thus, maintaining linearity is essential to ensure reliable current sensing and to prevent signal saturation or incorrect measurements.

\textbf{Search voltage scaling scheme.} To address this issue, we observed that those cells farther from the sensing block in the ML experience greater interconnect voltage drops, with the voltage drop exhibiting an approximately linear gradient along the ML towards the sensing block. To mitigate this non-ideality, we propose the use of a non-uniform search voltage across the search lines. Specifically, higher search voltages are applied to the cells farther from the sensing block, while lower search voltages are applied to cells closer to it. This approach generates higher mismatch voltages in distant cells, thereby increasing the gate voltage of the driver NMOS, resulting in a more balanced gate-to-source voltage across all the cells connected to ML. Fig.~\ref{fig:voltageScaling}(b) shows the ML current improvement with Hamming distance due to 4 level voltage scaling. This range of current is suitable for finding the least Hamming distance candidate using LTA block where minimum delta to distinguish is $0.2uA$ and the lowest current that can be sensed is in nA range\cite{donckers2000current}.

\begin{figure}
    \centering
    \vspace{-3pt}
    \includegraphics[width=.85\linewidth]{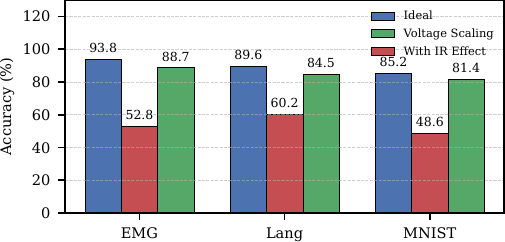}
    \vspace{-10pt}
    \caption{ The proposed voltage scaling scheme impact on performance}
    \vspace{-10pt}
    \label{fig:perfImprVoltageScaling}
\end{figure}

% \vspace{-5pt}
\section{Implementation and Evaluation Results}
\label{sec:implementation}

% \vspace{-1pt}
For our proposed design, we use ASAP $7nm$ PDK and a physics-based experimentally validated model for SOT layer and MTJ  \cite{narla2022design}. The MTJs diameter is of 45nm and $t_{ox}$ is 2nm which result in resistance for parallel and anti-parallel states to be $1.25M\Omega$ and $3.44 M\Omega$, respectively. 3.3nm thick AuPt layer has been used as the SOT channel, where the thickness is optimized for the minimum write energy based on spin drift-diffusion model~\cite{zhu2018highly}. Search voltage used is $1V$ where the write voltage is $0.8V$. We have designed 128$\times$128 SOT-CAM array and executed operations including multiplication, permutation, addition and search to extract latency, power and energy consumption using HSPICE. In SPICE simulations we have also considered interconnect parasitic extracted from the physical layout to ensure fair comparison. We have used Synopsys Design Compiler to synthesize our proposed HDC adder and compared it with a conventional HDC adder in the same ASAP $7nm$ technology node. To understand the operation-level improvement for similarity search, we have also implemented a dot product module in RTL and synthesized it. Similarly, a bit shifter and a bit-wise multiplier have been implemented and synthesized on the same technology node to compare permutation and multiplication respectively. Note that in all CMOS implementations, where computation is off-memory, there is additional delay associated with memory-to-compute unit communication, which is required for permutation, similarity search, and multiplication operations.
\begin{figure}[t!]
    \centering
    \includegraphics[width=.85\linewidth]{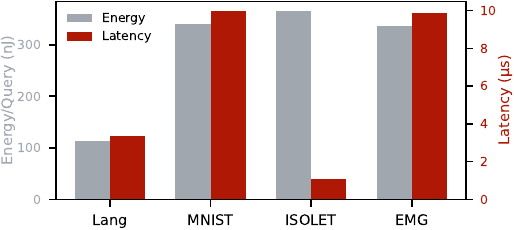}
    \vspace{-10pt}
    \caption{Energy per query and latency during inference for classifications.}
    \vspace{-10pt}
    \label{fig:clsPerfDs}
\end{figure}

\subsection{Validation of Design Choices}

\subsubsection{\textbf{Usage of binary HVs}}Fig. \ref{fig:Perf} reflects the design choices on 3 various applications: posture recognition from EMG signal, MNIST digit recognition and language recognition. We have measured the performance keeping all the HV's both binary and multibit during encoding and similarity search for comparison. In the binary form, only 1 and -1 are used to represent elements of a vector whereas in the multibit scheme a 16-bit integer is used for each element to allow large integer values. Our experimental results show that the binary design suffers only 3\%  accuracy degradation compared to the multibit design. However, we expect the binary representation to come with major benefits in terms of speed up in executing HDC operations and performing similarity search. Additionally, it enables energy and area efficient designs; hence, offering a good trade-off between accuracy and performance/energy.

\subsubsection{\textbf{Impact due to proposed permutation technique}} Fig.~\ref{fig:Perf} confirms that the proposed permutation technique where we have to drop batch wise elements during read have negligible impact on performance, enabling us to adopt an HDC algorithm driven strategy of dropping the first 8- or 16-bits during batch-wise reads of a given HV, thereby reducing latency and eliminating the need for an expensive bit shifter network in conventional designs.

\subsubsection{\textbf{Voltage scaling impact on performance}} To evaluate the impact of the proposed voltage scaling technique on performance, we measured the error introduced by ML current non-linearity and the sensitivity of the sensing modules. Due to irregular delta in current against Hamming distance, considering the IR effect shows a huge error as shown in Fig.~\ref{fig:perfImprVoltageScaling}. Voltage scaling improves the linearity of current with respect to Hamming distance, thereby reducing mispredictions and achieving accuracy comparable to the ideal scenario where current varies linearly with Hamming distance.

% \vspace{-5pt}
\subsection{Operation-wise Performance}
% \vspace{-1pt}
HyDra aims to exploit both operation-level and algorithm-level opportunities available in HDC. This enables fully in-memory binding, similarity search and permutation operations. To highlight operation-level improvements, we compared operation-level energy and latency with those of a custom circuit implementing multiplication, similarity search and permutation in the conventional form. We have designed a 2048-bit shifter, a bit-wise multitplier and a bipolar dot product module for similarity calculation. Note that, in HyDra during binding, similarity search and permutation, we do not require any memory to CPU communication which speeds up the system latency by an order of magnitude. Besides, we do not require any silicon area to accommodate the computation module. For the addition operation, we designed a conventional adder in system Verilog and compared it with our HDC-friendly implementation that contains only half adders and muxes.

Table.~\ref{Table:opComparison} summarizes the energy, area, and latency for our design and the standard custom implementation where the cycle time is $0.5ns$. We considered a $32Kb$ memory block of total $256kb$ which occupies a total area of $0.248mm^2$. To read 2048-bit HV, it takes 32 hops where one hop is $3ns$ or 6 cycles, resulting in total 192 cycles with a power consumption of $4.29mW$. For one HV read, the total energy consumption is $0.411nJ$. The results show that memory to CPU communication adds very high energy consumption. The values of the operation-wise energy dissipation are also less ($1.51\times$ for add, $6.19\times$ for permutation, $2.02\times$ for search) in HyDra compared to those of the all-CMOS implementation except for multiplication. This is because in multiplication we are considering write energy and latency in the target rows of the array which consumes significant portions of energy and latency. However, when considering the net energy per operation, including \( mem\_read \), the improvement achieved by HyDra is highly significant. Specifically, we observed reductions of \( 21.5\times \), \( 552.74\times \), \( 1.45\times \), and \( 282.57\times \) in energy consumption during addition, permutation, multiplication, and search operations, respectively.  

\begin{figure}[t!]
    \centering
    \includegraphics[width=.9\linewidth]{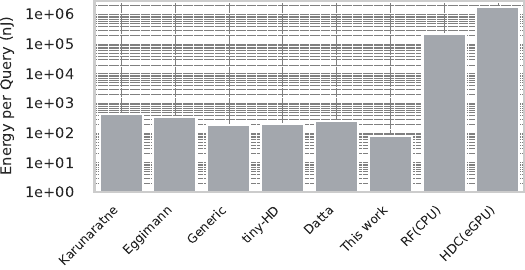}
    \vspace{-10pt}
    \caption{Comparison of HyDra's inference energy with other SOTA works.}
    % \vspace{-10pt}
    \label{fig:inferenceEnergy}
\end{figure}

\begin{figure}[t!]
    \centering
    % \vspace{-25pt}
    \includegraphics[width=.85\linewidth]{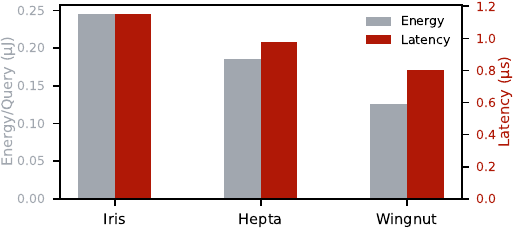}
    \vspace{-10pt}
    \caption{Energy and latency during clustering on different datasets.}
    \label{fig:clusteringELat}
    \vspace{-10pt}
\end{figure}

\begin{figure*}[tb]
    \centering
     \vspace{-5pt}
    \includegraphics[width=\textwidth]{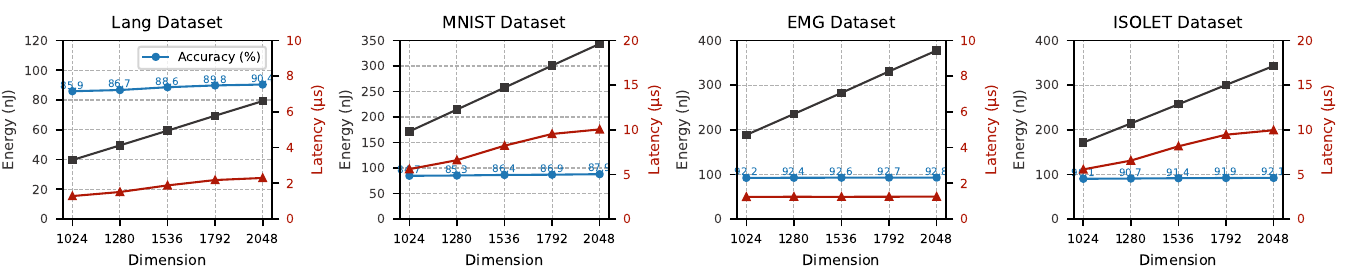}
    \vspace{-20pt}
    \caption{Energy, Latency and Accuracy vs HV dimensions which is achievable for reconfigurability of the CAM banks. Dataset-wise energy and latency profile can be improved by setting appropriate HV dimension without compromising accuracy.}
    \vspace{-5pt}
    \label{fig:dimVsEDA}
\end{figure*}

% \vspace{-5pt}
\subsection{Classification Performance}
% \vspace{-5pt}
As mentioned earlier, four types of datasets have been used during the experiments. Fig. \ref{fig:clsPerfDs} depicts the energy consumption and latency for each of the datasets. Among them ISOLET has the lowest latency as it does not require temporal encoding (i.e. no permutation). On the other hand, language recognition requires a comparatively small number of $ngram$ encoding because of its sequence length of average 100 letters. We have compared inference energy consumption with other SOTA work including Generic, tiny-HD and \cite{datta2019programmable,karunaratne2020memory,eggimann20215} along with random forest implemented in CPU and baseline HDC in eGPU.  Fig.~\ref{fig:inferenceEnergy} shows HyDra offers a $2.27\times$ reduction in energy consumption compared to Generic HDC accelerator where compared to CPU and GPU, it offers $2702\times$ and $23161\times$ reductions over RF in CPU and HDC baseline in eGPU, respectively.

% \vspace{-5pt}
\subsection{Clustering Evaluation}
% \vspace{-5pt}
We selected three clustering datasets (i.e., Iris, Hepta and Wingnut) to evaluate the proposed design. During clustering, data points are encoded once and stored in the CAM bank. After random initialization of HVs that represents K cluster centers, HVs corresponding to data points are assigned to clusters based on similarity. At each epoch, cluster center HVs are updated by summing the HVs of their assigned data points (using the adder block) and binarized. This process continues until the similarity distance between the old and updated cluster centers falls below a threshold. Note that during classification, encoding and similarity search operation counts are in the same order, while during clustering an extensive amount of similarity search is required before algorithm converges though encoding requires only once. The number of epochs differs across datasets, resulting in variations in latency and energy consumption, as shown in Fig.~\ref{fig:clusteringELat}. The illustration shows the energy and latency during execution of clustering for the datasets where Iris requires higher energy and latency due to higher epochs required to converge.

% \vspace{-5pt}
\subsection{Optimization through Reconfigurability}
% \vspace{-1pt}
HyDra supports HV dimension variability which offers flexibility to adapt different applications with optimized energy and latency profile. Fig.~\ref{fig:dimVsEDA} demonstrates the outcome of changing HV dimension on energy, latency and accuracy for various datasets. We observe large linear decrement trends in energy consumption per query and moderate latency reduction as well with respect to HV dimension reduction. However, the drop in accuracy is not identical for all datasets. This trend enables the configuration of HyDra for optimal energy efficiency and delay, while maintaining the expected performance, by adjusting the number of CAM banks.

\begin{figure}
    \centering
    \vspace{-15pt}
    \includegraphics[width=0.8\linewidth]{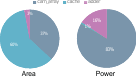}
    \vspace{-5pt}
    \caption{Area and Power breakdown of major blocks of HyDra.}
    \vspace{-10pt}
    \label{fig:piePA}
\end{figure}

% \vspace{-10pt}
\subsection{Power and Area}
% \vspace{-5pt}
Fig.~\ref{fig:piePA} depicts the area and power breakdown of HyDra for the major blocks. Cache for HV and CAM array occupy about $60\%$ and $37\%$ of total area, respectively. Cache has a larger area due to its cells, peripherals and large sensing blocks compared to CAM blocks. Adder block has $0.01185mm^2$ which is about $3\%$ of the total area. 
In power breakdown, we observe a higher portion of power coming from the CAM array as most of the computation is executed inside it while cache block is only used for temporary storage of class HV's. Due to size ($2048$ elements of $int16$ of HV is updated at once) of the adder block it consumes $16\%$ of the total power. 

% \vspace{-10pt}
\section{Conclusion}
\label{sec:conclusion}
% \vspace{-5pt}
HDC holds strong potential for a wide range of computational tasks. This work integrates emerging memory technology,  reconfigurable 5T-2MTJ SOT-CAM cells, for unified in-memory storage and computation of HDC operations. The proposed architecture supports key HDC functions across various hypervector dimensions. Through algorithm-hardware co-design, it enables in-memory permutation with a $6\times$ speedup over conventional designs, without additional hardware. A voltage scaling scheme mitigates IR drop, ensuring accurate similarity search via improved distance representation. Our design is general purpose and achieves a $2.27\times$ reduction in inference energy over SOTA HDC accelerator, and over three orders of magnitude lower energy than CPU and eGPUs. The system's idle power is negligible due to non-volatility. Efficiency can be further boosted by CAM bank reconfiguration. Beyond energy savings, the high density and non-volatility of the hardware yield superior area and energy efficiency, making it ideal for edge deployment. The system supports up to 300K queries per second with less than 3\% performance degradation, enabled by binary HV representation.

\bibliographystyle{IEEEtran}
\bibliography{references}

\end{document}